\documentclass[12pt]{iopart}
\usepackage{graphicx}
\usepackage{epsfig}

\def\apj{{\sl Astrophys.\ J. \ }}

\def\apjs{{\sl Astrophys.\ J.\ Supp. \ }}

\def\ijmpa{{\sl International\ J.\ Mod.\ Phys. \ A \ }}

\def\jcap{{\sl J.\ Cosm.\ Astroparticle\ Phys.  }}
\def\lrr{{\sl Liv.\ Rev. Rel. \ }}
\def\mnras{{\sl MNRAS \ }}
\def\n{{\sl Nature \ }}

\def\plb{{\sl Phys.\ Lett.\ B \ }}

\def\pr{{\sl Phys.\ Rep. \ }}
\def\prd{{\sl Phys.\ Rev.\ D \ }}
\def\prdrc{{\sl Phys.\ Rev.\ D \ Rap.\ Comm. }}

\def\rmp{{\sl Rev.\ Mod.\ Phys. \ }}

\newcommand{\gsim}{\,\lower2truept\hbox{${>\atop\hbox{\raise4truept\hbox{$\sim$}}}$}\,}
\newcommand{\pp}{~~~.}
\newcommand{\vv}{~~~,}

\def\etal{{\rm et al.$\,$}}

\newcommand{\be}{\begin{equation}}
\newcommand{\ee}{\end{equation}}
\newcommand{\bea}{\begin{eqnarray}}
\newcommand{\eea}{\end{eqnarray}}

\begin{document}

\title[Extended quintessence with an exponential coupling]{Extended quintessence with an exponential coupling}

\author{Valeria Pettorino$^{1}$, Carlo Baccigalupi$^{2}$, Gianpiero Mangano$^{1}$}
\address{$^{1}$ Universit\`{a} di Napoli {\it Federico II} and INFN, Sezione di
Napoli, \\ Complesso Universitario di Monte Sant'Angelo, Via Cintia,
I-80126 Napoli, Italy \\ $^{2}$SISSA/ISAS, Via Beirut 4, I-34014 Trieste,
Italy and
\\ INFN, Sezione di Trieste, Via Valerio 2, I-34127 Trieste, Italy}

\begin{abstract}
We study a class of extended quintessence cosmologies where the scalar field
playing the role of the dark energy is exponentially coupled to the Ricci scalar. \\
We find that the dynamics induced by the effective gravitational potential
in the Klein-Gordon equation dominates the motion of the field in the early
universe. The resulting ``$R-$boost" trajectory is characterized by a kinetic dark
energy density, given by $\left[3\rho_{mnr0}(1+z)\right]^2 \left[ 32
\rho_{r0} \, \omega_{JBD\, 0}\right]^{-1}$, where $\omega_{JBD\, 0}$, $\rho_{r0}$ and
$\rho_{mnr0}$ are calculated at present, and represent the Jordan Brans Dicke
parameter, the density of relativistic matter and of those species which
are non-relativistic at redshift $z$, respectively.
We show that such a trajectory represents an attractor, equivalent to a tracking
solution with equation of state $w=-1/3$,
providing a large basin of attraction for the initial dark energy density regardless
of the properties of the potential energy yielding acceleration today. We derive
the up to date constraints from Big Bang Nucleosynthesis (BBN) on the
present scenario, and we show that they are largely satisfied for interesting
trajectories of the dark energy scalar field in the early universe. \\
We compute the cosmological perturbation spectra in these cosmologies. For
a fixed value of $\omega_{JBD\, 0}$, the projection and Integrated Sachs-Wolfe
effects on the cosmic microwave background anisotropy are considerably larger in
the exponential case with respect to a quadratic non-minimal coupling, reflecting
the fact that the effective gravitational constant depends exponentially on the
dynamics of the field.
\end{abstract}

\maketitle

\section{Introduction}
\label{i}

According to the present cosmological observations, the universe
today is nearly geometrically flat, with an expansion rate of
about 72 km/sec/Mpc, and structures grown out of a primordial
linear spectrum of nearly Gaussian and scale invariant energy
density perturbations. About 5\% of the critical energy density is
made of baryons, while 20\% is thought to be composed by particles
interacting at most weakly with them (Cold Dark Matter, CDM). The
remaining 75\% is some sort of vacuum component, the dark energy,
with negative pressure acting as a repulsive gravitational force,
and responsible for a late time cosmic acceleration era. Several
independent observables support the picture above, {\it i.e.} type
Ia supernovae (hereafter SNIa
\cite{riess_etal_1998,perlmutter_etal_1999}), the Cosmic Microwave
Background (CMB) anisotropies (see \cite{bennett_etal_2003} and
references therein), the Large Scale Structure (LSS, see
\cite{tegmark_etal_2004} and references therein) and the Hubble
Space Telescope (HST \cite{freedman_etal_2001}).

While a dark matter component is generically predicted in the
super-symmetric extensions of the standard model for particle physics,
the dark energy component poses severe questions to the whole picture
(see \cite{padmanabhan_2003,peebles_ratra_2003} and references therein). Specifically,
a reliable dark energy candidate should explain why the present
amount is so small with respect to the fundamental scale
(a fine tuning of 123 orders of magnitude if compared to the Planck
scale), and why it is comparable with the critical density today
(coincidence). \\
The Cosmological Constant is affected by both these problems. As
an alternative, the concept of vacuum energy in cosmology has been
generalized in terms of a scalar field, named quintessence, which
for a broad class of potentials is able to converge to the present
regime starting from a wide set of initial conditions in the past,
thus alleviating the fine tuning (scaling or tracking
trajectories, \cite{liddle_scherrer_1999,steinhardt_etal_1999}).
The lack of a clear prediction for the dark energy scalar field in
any fundamental theory motivated the investigation, also
phenomenological, of its coupling with other cosmological
components, such as dark matter \cite{amendola_2004,MMP} or
gravity \cite{matarrese_etal_2004}.

This work belongs to the second class. The interest in connecting dark
energy and gravity is related to the original Brans-Dicke idea (see
\cite{fujii_maeda_2003} and references therein), and on the temptation to
explain the evidence for cosmic acceleration entirely in terms of a
modification of general relativity. Although such attempts were
unsuccessful so far, several interesting effects have been discovered in
these scenarios, generally called extended quintessence models,
concerning background evolution as well as cosmological perturbations
\cite{matarrese_etal_2004}, \cite{chiba_1999}-\cite{nojiri_etal_2004}.\\
A major achievement concerns the early universe dynamics of the
quintessence field in these scenarios. It was shown that the
species which are non-relativistic in the radiation dominated era
make the Ricci scalar diverging, activating an effective potential
in the Klein-Gordon equation coming from the non-minimal coupling
only ($R$-boost, \cite{baccigalupi_etal_2000}). This effect is
acquiring a crucial importance recently: as we stressed above, a
broad class of quintessence potentials allows for a large basin of
attraction in the early universe, alleviating the corresponding
fine tuning problem. On the other hand, the latter condition holds
only if the present dark energy equation of state $w$ is
sufficiently far from $-1$. As the observations are constraining
$w$ to be close to $-1$ (see {\it e.g.} \cite{tegmark_etal_2004})
the basin of attraction in the early universe shrinks, threatening
the basis of the whole quintessence picture \cite{bludman_2004}.
It was recently argued that in these conditions the coupling of
the dark energy with other entities may be relevant at early
times, possibly saving the existence of attractors for the initial
field dynamics \cite{matarrese_etal_2004}.

Despite the broad interest on this subject, most of the analysis so far was carried
out considering a quadratic coupling between the dark energy scalar
field and the Ricci scalar. In this paper we work out the first generalization of
those claims, providing a comprehensive analysis of cosmologies involving an
exponential coupling. The latter case may be relevant for string cosmology, where
the dilaton appears in an exponential in the fundamental lagrangian
\cite{gasperini_veneziano_2003}, although in the usual formulation the
coupling involves all terms, not just the Ricci scalar.

This paper is organized as follows. In Section II we describe the framework
of exponential couplings in extended quintessence cosmology. In Section III
we discuss the observational constraints on the coupling magnitude from
solar system experiments and Big Bang Nucleosynthesis. In Section IV we
study the $R$-boost trajectories of the dark energy scalar field. In
Section V we work out the most important effects on cosmological
perturbations. Finally, in Section VI we draw our conclusions.

\section{Exponential scalar-tensor couplings in cosmology}
\label{estcic}

A wide class of cosmological models can be described by the action \be
\label{action} S = \int d^4x \sqrt{- g}\left[ \frac{1}{2\kappa}
f(\phi,R) - \frac{1}{2} \omega(\phi)\phi^{;\mu}\phi_{;\mu} - V(\phi) +
{\cal{L}}_{\rm{fluid}}\right] \vv \ee where $g$ is the determinant of the
background metric, $R$ is the Ricci scalar, $\phi$ is a scalar field whose
kinetic energy and potential are specified by $\omega(\phi)$ and $V(\phi)$,
respectively. ${\cal{L}}_{\rm{fluid}}$ includes contributions from all
components different from $\phi$ and $\kappa=8\pi G_{*}$ represents the
bare gravitational constant \cite{esposito-farese_polarski_2001}. The usual
gravity term $R/16\pi G$ has been generalized by the function
$f(\phi,R)/2\kappa$, where $f(\phi,R)$ may describe a coupling between the
quintessence field $\phi$ and gravity, as well as a pure geometrical
modification of general relativity, featuring a non-linear dependence on
$R$. Note also that the gravitational constant appearing in the Lagrangian
and the one measured in Cavendish like experiments differ by corrections
being negligible in the limit $\omega_{JBD}\gg 1$
\cite{esposito-farese_polarski_2001}. The Lagrangian in (\ref{action}) has
been considered and analysed in its full generality in a cosmological
context including linear perturbations \cite{hwang_1991}; this framework
has been exploited in several works
\cite{matarrese_etal_2004,perrotta_etal_2000,baccigalupi_etal_2000,perrotta_baccigalupi_2002,
perrotta_etal_2004,caresia_etal_2004}, leading to the formulation of the
weak lensing theory \cite{acquaviva_etal_2004}.

In this work we consider a simple class of extended quintessence
models with exponential coupling \be \label{F_new} \omega(\phi)=1
\ \ ,\ \
\frac{f(\phi,R)}{2\kappa}=\frac{F(\phi)R}{2}=\frac{R}{16\pi
G}\exp\left[\frac{\xi}{m_{P}}\left(\phi-\phi_0\right) \right] \vv \ee where
$\xi$ is a dimensionless coupling, $\phi_0$ is the present value
for the $\phi$ field and $m_P$ the Planck mass $m_{P}=1/\sqrt{G}$
in natural units.
Note that the constant $\phi_{0}$ has been introduced to make explicit that
at present $F(\phi_{0})=1/8\pi G$. \\
The Jordan-Brans-Dicke parameter in this scenario is \be
\label{wJBD_def}
\omega_{JBD}=F \left(\frac{dF}{d \phi} \right)^{-2}=\frac{8\pi}{\xi^{2}}
\exp\left[-\frac{\xi}{m_{P}}\left(\phi-\phi_0\right)\right]\ .\ee
Note that at present $\omega_{JBD\, 0}=8\pi/\xi^{2}$.

We give now a brief overview of the relevant equations describing
the cosmological expansion assuming a Friedmann Robertson Walker
(FRW) background metric, since it is relevant in the following. In
Section V we describe the effects on the cosmological
perturbations; however, we do do not write down explicitely the
perturbation equations here, which can be found elsewhere (see
{\it e.g.}
\cite{hwang_1991,perrotta_baccigalupi_2002,acquaviva_etal_2004})
written for a generic function $F$. The most relevant differences
with respect to ordinary cosmologies are represented by the time
variation of $F$, which plays the role of a varying gravitational
constant in the Poisson equation. \\
For flat cosmologies, the line element can be written as $
ds^2 = a^2(\tau)(-\, d\tau^2 + \delta_{ij}dx^i dx^j)\,,
$
where $a(\tau)$ is the scale factor, $\tau$ represents the
conformal time variable, related to the cosmic time $t$ by the
transformation $dt = a(\tau)\,d\tau$. \\
The expansion and field dynamics are determined by the Friedmann
and Klein Gordon equations
\be
\label{FRWH2} {\cal H}^2 = {\left(\frac{\dot{a}}{a}\right)}^2 =
\frac{1}{3F}\left(a^2 \rho_{fluid} + \frac{1}{2}{\dot{\phi}}^2 +
a^2 V - 3{\cal{H}}{\dot{F}}\right)\ ,
\ee
\be
\label{KG}
\ddot{\phi} + 2{\cal{H}}\dot{\phi} =
\frac{a^2}{2}\frac{dF}{d\phi}R - a^2 V_{\phi} \ ,
\ee
where dot means derivative with respect to the conformal time $\tau$ and
$V_{\phi}$ is the derivative of the quintessence potential with respect to
$\phi$. For our analysis, it is relevant to write down the explicit form of
the Ricci scalar in terms of the cosmological content
\be
\label{ricci}
R=-\frac{1}{F}\left[-\rho_{fluid} + 3p_{fluid} +
\frac{\dot{\phi}^2}{a^2} - 4V + 3
\left(\frac{\ddot{F}}{a^2}+2\frac{{\cal{H}}\dot{F}}{a^2}\right)\right]\vv
\ee where $\rho_{fluid}$ and $p_{fluid}$ are the energy density
and pressure that need to be summed up over all possible
cosmological components but $\phi$. Note also that, as it was
shown in \cite{hwang_1991}, the usual conservation equations
$\dot{\rho}_i =-3{\cal{H}}(\rho_i + p_i)$, where $\rho_i$ and
$p_i$ are respectively the energy density and pressure of the
$i$-th component, hold for all species but $\phi$.

\section{Observational constraints}
\label{oe}

In this Section we briefly review the constraints on scalar-tensor
cosmology coming from the solar system physics \cite{will_2001}
and derive the up to date bounds from the BBN. \\
The time variation of the gravitational constant
is related to the coupling function $F$ by the expression \be \label{G_var_F}
\left|\frac{1}{G} \frac{d G}{d t}\right|=\left|\frac{1}{F} \frac{d F}{ d t}\right| \vv \ee
which does not take into account corrections due to the effective
gravitational constant in Cavendish like experiments which are small in the
limit $\omega_{JBD}\gg 1$ \cite{esposito-farese_polarski_2001}. Both local
laboratory and solar system experiments constrain the ratio in
(\ref{G_var_F}) to be less than $10^{-11}$ per year \cite{will_2001}. An
independent bound is given by the effects induced on photon trajectories by
a varying gravitational constant, which was recently greatly strengthened
by the Cassini solar system probe \cite{bertotti_etal_2003}: at two sigma,
the new constraint is
\be
\label{wjbd_constr}
\omega_{JBD\, 0}\ge 4{\times} 10^{4}\ .
\ee
It may not be straightforward to extend these limits, obtained on solar system
scales, to cosmology, as it was pointed out recently \cite{barrow_2004}.
Indeed we probe regions well within our Galaxy, which is a self-gravitating virialized
system: in physical theories where fundamental constants vary, the latter
may acquire local values which are different from their large scale,
cosmologically effective ones. For this reason, cosmology is likely to
become a source of constraints for the underlying theory of gravity
\cite{acquaviva_etal_2005}, in a complementary way with respect to the solar
system.

The BBN is indeed the only direct cosmological probe for the field value
in the radiation dominated era. As it is well known, the amount of
light nuclides produced when the photon temperature was in the range 0.01
$\div$ 10 MeV is rather sensitive to the value of the Hubble parameter
during that epoch, as well as to its time dependence. In particular the
$^4$He mass fraction $Y_p$ strongly depends on the freeze-out temperature
of weak processes which keep neutrons and protons in chemical equilibrium.
Changing the value of ${\cal H}$ affects the neutron to proton number
density ratio at the onset of the BBN, which is the key parameter entering the
final value of $Y_p$ and more weakly in the Deuterium abundance. For a fixed
baryon density parameter $\omega_b=\Omega_b h^2$ in the range suggested by
results of the Wilkinson Microwave Anisotropy Probe (WMAP),
$\omega_b=0.023 {\pm} 0.001$ \cite{spergel_etal_2003} both nuclei
yields are in fact monotonically increasing functions of ${\cal H}$. For
recent reviews on BBN see \cite{cuoco_etal_2003, serpico_2004,
cyburt_2004}.

As we read from (\ref{FRWH2}) the introduction of the field $\phi$
changes the value of ${\cal H}$ in two different ways. First of
all there is a shift in the effective gravitational constant by a
factor $\exp[-\xi (\phi- \phi_0)/m_P)]$. Second, it gives an extra
contribution to ${\cal H}$. By using the coupling defined in
(\ref{F_new}), we can solve (\ref{FRWH2}) with respect to ${\cal
H}$, getting \be \fl {\mathcal H} = \sqrt{\frac{8\pi G}{3}}
\left[\sqrt{a^2\left(\rho_{fluid}+\rho_\phi \right) \exp\left(-
\sqrt{\frac{8\pi}{\omega_{JBD\, 0}}}\frac{\phi
-\phi_0}{m_{P}}\right)+\frac{3 \dot{\phi}^{2}}{4 \omega_{JBD\,0}}}
- \sqrt{\frac{3}{4 \omega_{JBD\,0}}}\,\dot{\phi}\,\right]\ .
\label{hubble1} \ee In the following we shall assume that the
Universe is indeed radiation dominated during the BBN epoch, so
that $\rho_\phi$ can be neglected with respect to $\rho_{fluid}$
which receives contributions from photons, neutrinos and
$e^{\pm}$. Moreover the last two terms in Equation (\ref{hubble1})
are suppressed by $1/\sqrt{\omega_{JBD\,0}}$. In this case the
squared Hubble parameter takes the simple form
\begin{equation}
H^2 = \frac{{\mathcal H}^2}{a^2} = \frac{8 \pi G}{3}\,
\exp\left[-\frac{\xi (\phi -\phi_{0})}{m_P}\right] \,\rho_{fluid} \,\,\,.
\label{hubble2}
\end{equation}
In this limit we see that by comparing the theoretical values for
the light nuclei abundances with the corresponding experimental
determinations we can constrain the value of the effective
gravitational constant during the BBN epoch, {\it i.e.} the
quantity $\xi (\phi -\phi_{0})/m_P$.

Before discussing this issue let us briefly summarize the present
knowledge of the BBN both from the experimental and theoretical
points of view, see {\it e.g.} \cite{serpico_2004} for a review.
The most accurate measurement of primordial Deuterium number
density, normalized to Hydrogen, $X_D$, is obtained from DI/HI
column ratio in QSO absorption systems at high redshifts. The most
recent estimate \cite{kirkman_etal_2003} gives
\begin{equation}
X_D^{exp}=(2.78^{+0.44}_{-0.38}) {\cdot} 10^{-5} \,\,\,,
\label{deut}
\end{equation}
and is in good agreement with the theoretical expectation for
a standard scenario with three active neutrinos and a baryon
density given by the WMAP result \cite{serpico_2004}; the latter is
\begin{equation}
X_D^{th}=(2.44^{+0.19}_{-0.17}) {\cdot} 10^{-5} \,\,\,,
\label{deutth}
\end{equation}
where the (1$\sigma$) theoretical uncertainty accounts for both
the propagated error due to the several rates entering the BBN
nuclear reaction network as well as the 5\% uncertainty on
$\omega_b$. On the other hand the $^4$He mass fraction $Y_p$
obtained by extrapolating to zero the metallicity measurements
performed in dwarf irregular and blue compact galaxies is still
controversial and possibly affected by large systematics. There
are two different determinations
\cite{fields_olive_1998,izotov_thuan_2004} which are only
compatible by invoking the large systematic uncertainty quoted in
\cite{fields_olive_1998}
\begin{eqnarray}
Y_p &=& 0.238 {\pm}(0.002)_{stat}{\pm}(0.005)_{sys} \,\,\,,\\
Y_p &=& 0.2421 {\pm}(0.0021)_{stat}\,\,\,.
\end{eqnarray}
Both results are significantly lower than the theoretical estimate
\cite{serpico_2004}
\begin{equation}
Y_p^{th} = 0.2481 {\pm} 0.0004 \,\,\,,
\end{equation}
where again we use $\omega_b=0.023$ and the small error is due to the
uncertainty on the baryon density and, to a less extent, to the error on
experimental determination of the neutron lifetime. As in
\cite{serpico_2004} we use in our analysis a more conservative estimate for
the experimental $^4$He abundance obtained by using the results of
\cite{olive_skillman_2004}
\begin{equation}
Y_p^{exp}=0.245 {\pm} 0.007 \,\,\,. \label{exphe}
\end{equation}
We finally recall that the Spite plateau value for $^7$Li found in PopII
dwarf halo stars is smaller by a factor 2 $\div$ 3 with respect to the
standard BBN prediction \cite{serpico_2004, cyburt_2004}. In view of this
problem, suggesting a large depletion mechanism of primordial $^7$Li, we do
not consider this nuclide in the following. \\
To bound the value of the quintessence field we construct the
likelihood function \begin{equation} {\cal L}(\xi \phi)\propto
e^{-\chi^2[\xi (\phi-\phi_{0})/m_P]/2}\,\,\,,
\end{equation}
with
\begin{eqnarray}
&& \fl \chi^2\left[\xi (\phi -\phi_{0})/m_P\right]= \nonumber
\\ && \fl \sum_{i,j={\rm D},^4{\rm He}} \left[X_i^{th}[\xi(\phi-
\phi_{0})/m_P]-X_i^{exp}\right]W_{ij}\left[X_j^{th}[\xi (\phi
-\phi_{0})/m_P]-X_j^{exp}\right] \,\,\,.
\end{eqnarray}
The proportionality constant can be obtained by requiring normalization to
unity  of the integral of ${\cal L}$, and with $W_{ij}$ we denote the
inverse covariance matrix
\begin{equation} W_{ij}=[\sigma_{ij,th}^2+\sigma_{i,exp}^2\delta_{ij}]^{-1}
\,\,\,,
\end{equation} where $\sigma_{i,exp}$ is the uncertainty in
the experimental determination of nuclide abundance $X_i$ and
$\sigma_{ij,th}^2$ the theoretical error matrix. We also consider
the two likelihood functions for each of the two nuclei to show
how at present the Deuterium and $^4$He can separately constrain
the value of the effective gravitational constant at the BBN epoch.

It is important to stress that we consider the simplest case of a
constant value of $\phi$ during the whole BBN phase. This is
justified a posteriori by the discussion in the next Section;
although the dynamics of the field in the radiation dominated era
is cosmologically relevant, it is too small to provide any change
of $\phi$ during BBN which is significant compared with the bounds
we derive here. It is also worth mentioning that in view of the
possible systematics affecting mainly the experimental
determination of $^4$He it is unfortunately impossible at present
to use BBN to get detailed constraints on the time evolution of
the quintessence field during this phase, while it can only
provide a conservative bound on the largest (or smaller) values
attainable by $\xi (\phi-\phi_0)/m_P$ during the nuclei formation
era.

\begin{figure}[!hbtp]
\begin{center}
\epsfig{file=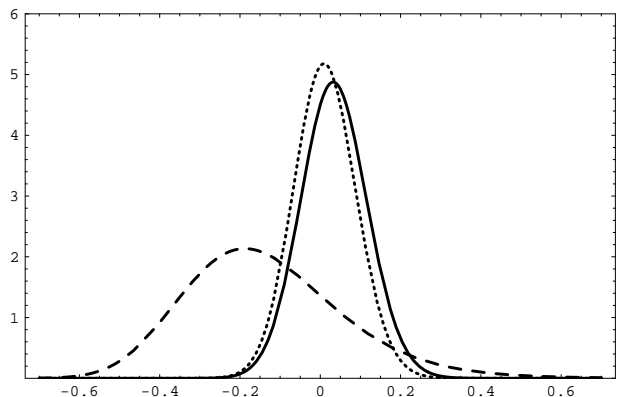,width=8.5truecm} \caption{The
behavior of the likelihood functions versus the parameter $\xi
(\phi-\phi_0)/m_P$ for $\omega_b=0.023$. The three curves refer to
the D (dashed curve), $^4$He (solid) and the combined D+$^4$He
(dotted) analysis discussed in the text. }\label{likeli}
\end{center}
\end{figure}

Our results are summarized in Figure \ref{likeli}, where we show the
likelihood contours obtained using Deuterium, $^4$He and finally, combining
the two nuclei. As expected the $^4$He mass fraction gives the most
stringent constraint since it is much more sensitive to the value of the
Hubble parameter with respect to Deuterium. In particular using $^4$He only
we get
\begin{equation}
-0.13 \leq \xi (\phi -\phi_{0})/m_P \leq 0.20 \,\,\,, \,\,\,\,95\% \,C.L. \label{bbn1}
\end{equation}
while adding the information on D abundance does not change this
result significantly:
\begin{equation}
-0.14 \leq \xi (\phi -\phi_{0})/m_P \leq 0.17 \,\,\,, \,\,\,\,95 \%\,C.L.\ .
\label{bbn2}
\end{equation}
We see how a clear understanding of the role of systematics in the
$Y_p$ measurements would have a large impact on further
constraints on the value of the effective gravitational constant, at
least in the minimal BBN scenario we are here considering, with no
other extra parameters, such as extra relativistic degrees of
freedom.

We close this Section by recalling that for a constant value of
$\phi$ during the BBN epoch, its effect in renormalizing the
gravitational constant can be also conveniently recasted in terms
of the effective number of relativistic species contributing to
the total energy density, as it was noticed in earlier works
\cite{perrotta_etal_2000,chen_etal_2001}. We parameterize the
value of the Hubble parameter at the onset of BBN and before the
$e^{\pm}$ annihilation phase in terms of an excess (or deficit) of
relativistic degrees of freedom $\Delta N$
\begin{equation}
H^2= \frac{8 \pi G}{3} \left(\frac{g_*}{2}+\frac{7}{8} \Delta N
\right) \rho_\gamma\ ,
\end{equation}
where $\rho_\gamma$ is the photon energy density and $g_*=10.75$
the standard value obtained by summing over photons, $e^{\pm}$ and
neutrinos. By using (\ref{hubble2}) it is straightforward to get the
simple relation
\begin{equation}
\xi \frac{\phi -\phi_{0}}{m_P} = - \log \left(1+\frac{7}{43} \Delta N
\right)\ .
\end{equation}
This relation can be used to recast the bounds on the effective
number of neutrinos, which are routinely used in the literature,
in terms of a constraint on $\xi (\phi -\phi_{0})/m_P$ for the
exponential coupling considered in this paper. As we show in the
next Section, interesting values of the field at Nucleosynthesis
largely satisfy the bounds represented in Figure \ref{likeli}.

\section{$R$-boost}
\label{rb}

In this Section we derive the dark energy dynamics before
the onset of acceleration. The central point is that
extended quintessence scenarios may possess attractor trajectories
in the early universe, generated only by the non-minimal coupling, as
it was pointed out recently \cite{matarrese_etal_2004}; that allows
to remove the fine tuning in the early universe, even if the potential
is constrained to be almost flat in order to be consistent with the observations;
the latter occurrence indeed limits the capability of minimally coupled quintessence
models to avoid the fine tuning problem \cite{bludman_2004}.
The attractors in minimally coupled quintessence models were called scaling
\cite{liddle_scherrer_1999} or tracking \cite{steinhardt_etal_1999}
solutions; the corresponding trajectories in extended quintessence coming
from the non-minimal coupling are called $R$-boost solutions
\cite{baccigalupi_etal_2000}. In the following we derive the $R-$boost
for exponential coupling. We shall write
an analytic expression for that, both in matter dominated
and radiation dominated eras, which is manifestly independent on
the initial conditions; that represents an important new aspect with
respect to earlier works \cite{baccigalupi_etal_2000}, in which the
initial field value appeared explicitely in the $R-$boost energy density.
Moreover, exploiting a perturbative analysis \cite{liddle_scherrer_1999},
we demonstrate that such solution represents
an attractor. Also, we shall see that the energy density
corresponding to the $R-$boost depends on the value of $\omega_{JBD}$.

We shall compare what we find here with previous models of
extended quintessence, based on a quadratic coupling between the
dark energy and the Ricci scalar, and with a case of minimally
coupled quintessence (QCDM). In order to satisfy the constraint
(\ref{wjbd_constr}) the constant $\xi$ is obtained through
(\ref{wJBD_def}) by fixing $\omega_{JBD0}=10^5$. The remaining
cosmological parameters are chosen consistently with the
concordance model (see {\it e.g.} \cite{spergel_etal_2003}). The
present dark energy density is $73\%$ of the critical density,
$\Omega_{\phi} = 0.73$, with a Cold Dark Matter contribution of
$\Omega_{CDM}= 0.226$, three families of massless neutrinos,
baryon content $\Omega_b = 0.044$ and Hubble constant $H_0 = 72$
Km/sec/Mpc. The quintessence potential is an inverse power law \be
\label{potential} V(\phi) = \frac{M^{4+\alpha}}{\phi^\alpha} \ ,
\ee with $\alpha$ close to zero in order to have $w_{\phi}$ close
to $-1$ at present (see {\it e.g.} \cite{baccigalupi_etal_2002}
and references therein). We choose the initial conditions at
$1+z=10^{9}$; in typical runs, with $w_{\phi}=-0.9$ at present,
the value of the field today is $\phi_{0}=0.35m_{P}$; the dynamics
induced by the potential $V$ and the $R-$boost makes
$\phi\ll\phi_{0}$ at early stages; the condition on $\omega_{JBD\,
0}$ sets the value of the coupling constant $\xi
=1.6{\times}10^{-2}$. As a consequence, the BBN bounds
(\ref{bbn1}) or (\ref{bbn2}) are largely satisfied.

\subsection{Radiation dominated era (RDE)}
\label{radiation_era}

\begin{figure}[!hbtp]
\begin{center}
\epsfig{file=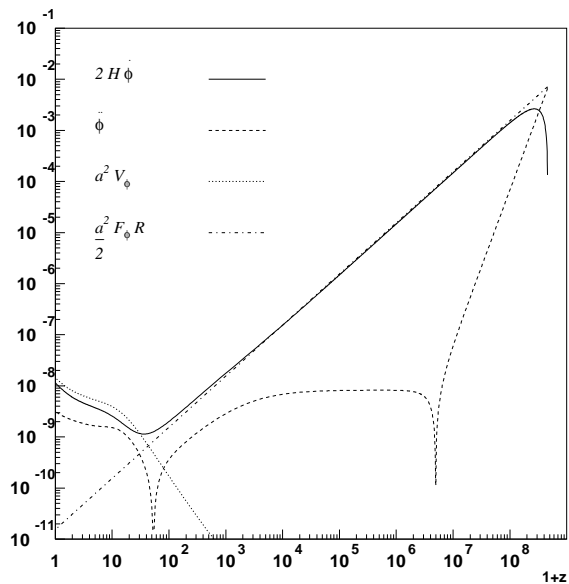,width=8.5truecm}
\caption{Absolute values of the four terms in the Klein Gordon equation
(\ref{KG_exp2}) for an exponential coupling.}
\label{fig_KG}
\end{center}
\end{figure}

The crucial point is that in (\ref{ricci}) the term
$\rho_{fluid}-3p_{fluid}$ gets no contribution from radiation but,
as long as one or more sub-dominant non-relativistic species
exists, the Ricci scalar gets a non-zero contribution that
diverges as $1/a^3$, thus becoming increasingly relevant at early
times \be \label{ricci_approx} R
\simeq\frac{1}{F}\frac{\rho_{mnr0}}{a^3} \, \, \, \, {\rm for} \, \,
a\rightarrow 0 \pp \ee The parameter $\rho_{mnr0}$ above indicates
the energy density of the species which are non-relativistic at
the time in which the dynamics is considered. As a consequence,
the term proportional to $R$ in the Klein Gordon equation acts as
a new source on the quintessence field besides
the true potential $V(\phi)$; the resulting motion is the $R-$boost. \\
With our choice of the coupling (\ref{F_new}) the Klein Gordon
equation (\ref{KG}) for the field can be written as \be
\label{KG_exp1} \ddot{\phi} + 2{\cal{H}}\dot{\phi} - \frac{1}{2}
\frac{\xi}{m_{P}}\frac{\rho_{mnr0}}{a} + a^2 V_{\phi} = 0 \vv \ee
where we have used (\ref{ricci}) to write the Ricci scalar in
terms of the total energy density, neglecting the remaining terms
in (\ref{ricci}) since they get negligible for $a$ sufficiently small.
\begin{figure}[!hbtp]
\begin{center}
\epsfig{file=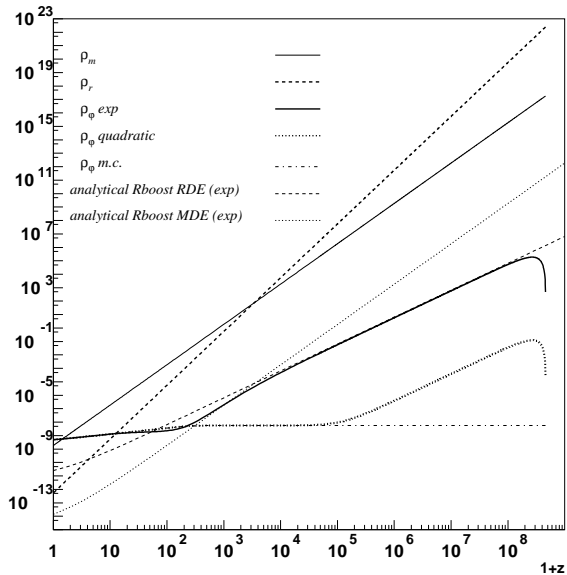,width=8.5truecm}
\caption{Energy density of matter (solid), radiation (heavy dashed) and
$\phi$ for minimal (dot dashed), quadratic (heavy dotted) and exponential
(heavy solid) coupling. Also shown the analytical behavior of the R-boost
during radiation dominated era (dashed) and matter dominated era (dotted).}
\label{rho}
\end{center}
\end{figure}
For our purposes here we neglect the departure from general
relativity in the RDE, {\it i.e.} we assume $F\simeq 1/8\pi G$ and
we ignore all the contributions in the Friedmann equation except
radiation. The above assumption is justified because we aim at
deriving the $R-$boost dynamics at first order in
$1/\omega_{JBD}$: a general analysis would yield corrections of
higher order in $1/\omega_{JBD}$ in the effective gravitational
potential appearing in the Klein Gordon equation. The expansion is
given by \be \label{a_rde} a \simeq C \tau \ , \ee where $C =
\sqrt{8 \pi G \rho_{r0}/3}$ and $\rho_{r0}$ represents the
cosmological radiation density calculated at present; since
${\cal{H}}\sim 1/\tau$, the equation of motion for $\phi$ reduces
to \be \label{KG_exp2} \ddot{\phi} + \frac{2}{\tau}\dot{\phi} -
\frac{1}{2}\frac{\xi}{m_{P}}\frac{\rho_{mnr0}}{C\tau}= 0 \vv \ee
which is solved by \be \label{KG_exp3_rboost} \phi(\tau ) =
\frac{\xi}{4}\rho_{mnr0}\sqrt{\frac{3}{8 \pi
{\rho_{r0}}}}(\tau-\tau_{beg}) + \phi_{beg} \vv\ee where
$\phi_{beg}$ is the initial condition for $\phi$. This solution
corresponds to the slow roll phase which starts when the
cosmological friction and effective gravitational potential
effects in the equation of motion are comparable, yielding the
$R-$boost equation \be \label{slowroll_rde} 2{\cal{H}}\dot{\phi}
\simeq \frac{a^2 R }{2} \frac{d F}{d \phi} \pp \ee In this phase,
the energy of the quintessence field is dominated by its kinetic
contribution \be \label{kin_rde} \frac{1}{2}\left( \frac{d \phi}{d
t}\right)^2 = \frac{3}{32}\frac{{\rho_{mnr0}}^2} {\rho_{r0}}
\frac{1}{\omega_{JBD\, 0}} (1+z)^2 \pp \ee from which it is
possible to derive that the R-boost solution is equivalent to a
tracking one with equation of state $-1/3$. It is interesting to
note that at first order in $1/\omega_{JBD}$ the $R-$boost energy
density is related only to the present value of the
Jordan-Brans-Dicke parameter. In Figure \ref{rho} we have plotted
the energy density of the various cosmological components: besides
matter and radiation we can observe the behavior of $\phi$ for
minimal and extended quintessence, starting from zero initial
kinetic energy. In the minimal coupling case the field behaves
nearly as a cosmological constant until the true potential $V$
starts to be relevant. In the extended quintessence case, instead,
the field accelerates and soon enters the $R-$boost phase. In
Figure \ref{fig_KG} we show the absolute values of the four terms
in the Klein Gordon equation (\ref{KG_exp2}). The potential term
starts to be dominant only for $z \leq 10^2$; the friction term is
zero at the beginning and then increases, joining the $R-$boost.
The quintessence field accelerates until the sign inversion in
$\ddot{\phi}$ occurs; then the field accelerates again for $z\sim
10^2$ when the potential $V$ starts to have a relevant effect. The
timing of the different phases of the trajectory depends on the
details we have fixed, but the general behavior is model
independent.

\subsection{Matter dominated era (MDE)}
\label{matter_era}

In the matter dominated era, equation (\ref{slowroll_rde}) still
holds but the expansion parameter has a different behavior with
time \be \label{a_mde}
 a = \frac{2}{3} \pi G \rho_{mnr0} {\tau}^2\pp
\ee As in the RDE case, we neglect the departure from general
relativity to get the $R-$boost dynamics at the lowest order in
$1/\omega_{JBD}$. The $R-$boost equation is \be
\label{slowroll_mde} \dot{\phi} = \frac{3}{16 \pi}
\frac{\xi}{\tau} m_P \vv \ee which is solved by \be \label{KG_mde}
 \phi = \frac{3}{16 \pi} \xi m_{P} \log{\frac{\tau}{\tau_{beg}}} +
\phi_{beg} \vv \ee also shown in Figure \ref{rho}. As a
consequence, the behavior of the kinetic energy of the field changes, too
\be \label{kin_mde} \frac{1}{2} \left(\frac{d \phi}{d t}\right)^2 =
\frac{3}{32}\frac{1}{\omega_{JBD\,
0}}\rho_{mnr0}(1+z)^3\pp \ee If we now look at Figure \ref{rho} again, we
notice that the $R-$boost has a bigger effect on $\rho$ in the exponential
case with respect to the quadratic coupling, also shown in the figure. In
both cases $\phi$ receives a strong kick, which determines a major change
in the dynamics with respect to the minimally coupled scenario. However,
for an exponential coupling the path of the field departs from the standard
one earlier with respect to the case of a quadratic coupling, reflecting
the fact that the exponential coupling enhances the departure from general
relativity as it depends exponentially on the field dynamics. We shall come
back to this issue in the next Section.

\subsection{Stability}
\label{stability}

Following the method exploited in \cite{liddle_scherrer_1999}, we
now show that the $R-$boost solution above is an attractor. We
look for scaling solutions of the Klein Gordon equation, {\it
i.e.} solutions in which the energy density of quintessence field
scales as a power of the scale factor \be\label{scaling_rde}
\left(\frac{d \phi}{d t}\right)^2 \propto a^{-2} \vv \ee in the RDE and
\be\label{scaling_mde} \left(\frac{d \phi}{d t}\right)^2 \propto a^{-3} \vv \ee in the
MDE, as found in (\ref{kin_rde}, \ref{kin_mde}). The behavior of
the background energy density is also given by a power of the
scale factor. In particular \be \label{energy_density_rad} \rho_r
\propto a^{-4} \,\,\, , \,\,\,\,\,\,\,\,\,\,\,\, \rho_{m} \propto
a^{-3} \vv \ee for radiation and non-relativistic matter respectively. As
far as $a(t)$ is concerned, neglecting again the corrections of the order
$1/\omega_{JBD}$ induced by $1/F$ in front of the right hand side of the
Friedmann equation (\ref{FRWH2}) as well as the other dark energy terms,
one has \be
\label{a_t_RDE}a^{RDE} \propto t^{1/2} \vv \ee \be
\label{a_t_MDE}a^{MDE} \propto t^{2/3} \ . \ee The corresponding
time dependence of the field $\phi$ is given by \be
\label{phi_e_RDE} {\phi_e}^{RDE} = \phi_*
\left(\frac{t}{t_*}\right)^{\frac{1}{2}} \vv \ee \be
\label{phi_e_MDE} {\phi_e}^{MDE} = \phi_*\log
{\left(\frac{t}{t_*}\right)} \vv \ee where the subscript `$*$'
stands for a given reference time and the subscript `e'
reminds us that this is the $R-$ boost exact solution. \\
We verify that the $R-$boost solution is an attractor by rewriting the
Klein Gordon equation (\ref{KG}), with the change of variables given by \be
\label{change_u} u = \frac{\phi}{\phi_e} \vv \ee and \be
\label{change_t} \tilde{\tau} = \log\left(\frac{t}{t_*}\right) \pp \ee
During RDE, we obtain
\be \label{KG_u_rde}
2u''+3u'+u-1 = 0 \vv \ee where the prime stands for the derivative with
respect to $\tilde{\tau}$. As we immediately see, this equation admits a
critical trajectory for $u = 1$ and $u' = 0$, corresponding to the
$R-$boost. Linearizing equation (\ref{KG_u_rde}) by choosing small
exponential perturbations around the critical point ($u = 1 +
e^{\lambda\tau}$) and solving for the eigenvalues $\lambda_{1,2}$ we get
\be\label{lambda_eig} \lambda_1 = -1 \, \, \, ; \, \, \, \lambda_2
= -\frac{1}{2} \pp \ee
The fact that the eigenvalues are real and negative shows that the generic
perturbation will be suppressed with time, thus flattening the trajectory
on the critical one $\phi
=\phi_{e}$. \\ Similarly, during the MDE, the Klein Gordon equation
(\ref{KG}) can be rewritten as
\be \label{KG_u_mde} u''+u' = 0 \ , \ee
which again admits $u=1$ as a solution. In this case, by substituting $u =
u + e^{\lambda\tau}$ in the equation and solving for the $\lambda$
eigenvalues, we find the two values $\lambda_1 = 0$ and $\lambda_2 = -1$,
independently of the value of $u$. As a consequence, the generic solution
of the Klein Gordon equation will tend to $u=$ constant. The constant is
fixed to be equal to $1$ by the initial conditions for the MDE,
corresponding to the final regime in the RDE, where the trajectory joined
the $R-$boost solution represented by $\phi_{e}$.

Summarizing, the simple choice of an exponential as a coupling between dark
energy and Ricci scalar in the Lagrangian leads to the existence of an
attractive $R-$boost solution in both the matter and radiation dominated
eras.

\begin{figure}[!hbtp]
\begin{center}
\epsfig{file=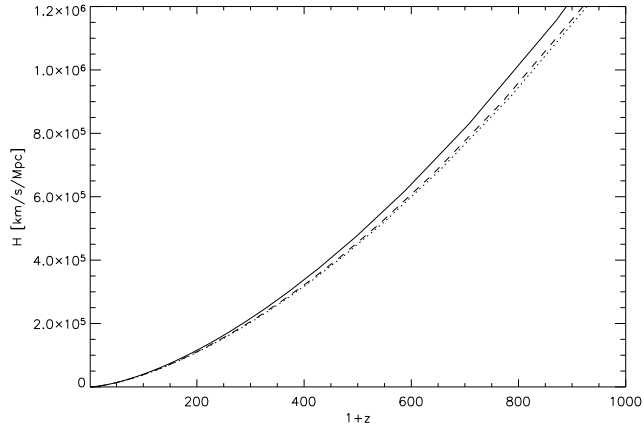,width=8.5truecm} \caption{The behavior of $H$ as
a function of the redshift in three cosmological models, extended
quintessence with exponential coupling (solid), quadratic coupling
(dashed) and QCDM (dotted).} \label{h}
\end{center}
\end{figure}

\section{Effects on cosmological perturbations} \label{eocp}

In this section we analyze the effect that the exponential coupling in
(\ref{action}) has on the spectra of cosmological perturbations. We find
that these effects are qualitatively consistent with those corresponding to
a quadratic coupling, but quantitatively different. For our purposes, an
analysis based on the CMB power spectrum only is sufficient. The scalar
perturbations are Gaussian and described by a scalar power spectrum with
spectral index $n=0.96$, and no tensors, consistently with the cosmological
concordance model, see \cite{spergel_etal_2003} and references therein. \\
Since the constant $\xi$ is chosen to be positive and the field $\phi$ is
smaller in the past with respect to $\phi_0$, $F < 1/8\pi G$ in the past.
As a consequence, our model describes a cosmology in which the
gravitational constant (and thus the Hubble parameter) is higher in the
past than in the QCDM case. It follows that for a fixed $\omega_{JBD\, 0}$,
we expect a larger amplitude of the effects induced by the behavior of $F$
for the exponential case with respect to the quadratic coupling: indeed,
the exponential is sensitive to the field dynamics also at the linear
level, which dominates for small values of $\phi /\phi_{0}-1$. This can be
seen in Figure \ref{h}, where we plot $H(z)$ for three cosmological models:
two of them represent extended quintessence with the same value of
$\omega_{JBD}$ at present, but featuring an exponential and quadratic
coupling, while the third one is the QCDM case. For the first two cases, we
have set $\omega_{JBD\, 0}=50$; we stress that we have chosen this small
value with respect to the existing bounds from solar system
\cite{bertotti_etal_2003}, in order to highlight the differences in the
models considered. The value of $H_{0}$ is the same, but $H(z)$ are quite
different functions in the three cases, in particular for the exponential
case with respect to the QCDM cosmology.

\begin{figure}[!hbtp]
\begin{center}
\epsfig{file=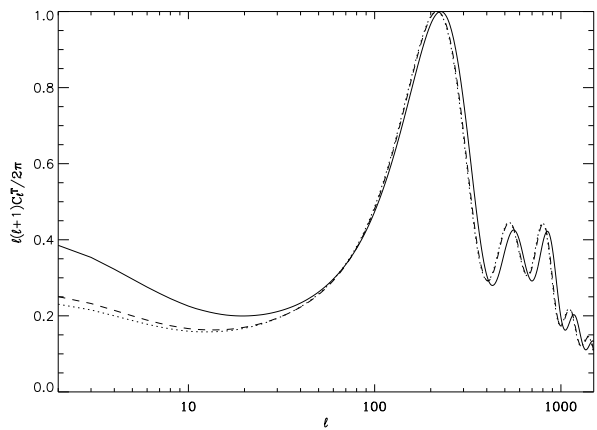,width=8.5truecm} \caption{CMB angular total
intensity power spectra for a QCDM cosmology (dotted), quadratic
(dashed) and exponential coupling extended quintessence (solid)
with $\omega_{JBD}=50$. The spectra are in arbitrary units,
normalized to 1 at the first acoustic peak in total intensity.}
\label{fig_clt}
\end{center}
\end{figure}

\begin{figure}[!hbtp]
\begin{center}
\epsfig{file=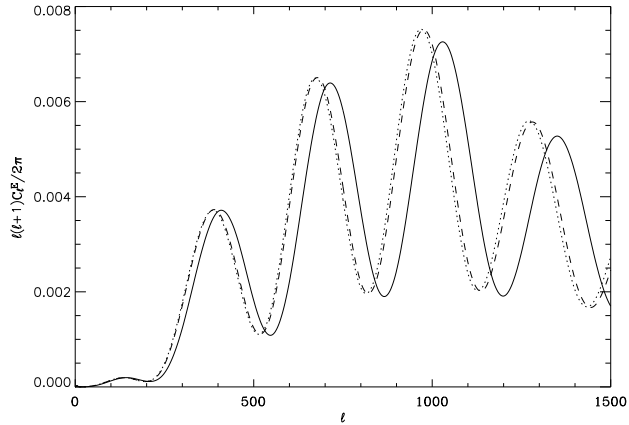,width=8.5truecm}
\caption{CMB angular polarisation power spectra for a QCDM
cosmology (dotted), quadratic (dashed) and exponential coupling
extended quintessence (solid) with $\omega_{JBD}=50$.
The spectra are in arbitrary units, normalized to 1 at the first
acoustic peak.}
\label{fig_cle}
\end{center}
\end{figure}

In Figures \ref{fig_clt} and \ref{fig_cle} we plot the total intensity and
polarisation power spectra of anisotropies for the models considered in
Figure \ref{h}. Two features are immediately evident. First the different
amplitude in the tail at low multipoles, and second a projection difference
in the location of the acoustic peaks. The spectra have been normalized to
the amplitude of the first peak, which roughly corresponds to fixing the
amplitude of the signal at last scattering. Both the effects come from the
different behavior of the coupling $F$ with redshift. The projection
feature simply follows from the difference in the curves in Figure \ref{h}.
Indeed, smaller (higher) values of $H^{-1}$ project the CMB power spectrum
onto smaller (larger) angular scales in the sky. The power at low
multipoles is modified through the Integrated Sachs-Wolfe (ISW) effect,
which is sensitive to the change of the cosmic equation of state at low
redshift (see \cite{hu_1998} and references therein). The larger is that
change, the larger is the ISW power.

Summarizing, the different form of the coupling function $F$ determines a relevant
difference of amplitude in the effects induced by extended quintessence models
on the cosmological perturbations. Larger departures from general relativity in
the dynamics of the field in the early universe correspond to larger effects on
the perturbations. The latter aspect has to be taken into account when constraining
different theories on the basis of the present observations.

\section{Conclusions}
\label{c}

We made a general analysis of quintessence cosmologies where the dark
energy scalar field is exponentially coupled with the Ricci scalar.
With respect to the case of a non-minimal quadratic coupling, we find
relevant new results concerning both background expansion and
perturbations.

The dynamics of the field in the early universe is dominated by the
effective potential in the Klein-Gordon equation, entirely due to the
non-minimal interaction. The presence of non-relativistic species, although
sub-dominant with respect to radiation, makes the behavior of the Ricci
scalar actually diverging in the early universe, determining the dynamics
of the field. The whole trajectory, the $R$-boost, corresponds to the slow
roll of the field on the effective gravitational potential. \\
Interestingly, in the present scenario the $R$-boost looses any sensitivity
on the initial conditions of the field, while in the case of a quadratic
coupling the latter enters into the expression for the energy density along
the trajectory. The $R-$boost solution is then equivalent to a tracking one
with equation of state $-1/3$, and energy density
$\left[3\rho_{mnr0}(1+z)\right]^2 \left[ 32
\rho_{r0} \, \omega_{JBD\, 0}\right]^{-1}$, where $\rho_{r0}$ is the radiation
energy density today, $\rho_{mnr0}$ the present energy density of the
species which are non-relativisitc at redshift $z$, and $\omega_{JBD\, 0}$ is
the Jordan-Brans-Dicke parameter, also calculated at present. This aspect
is extremely relevant in dark energy cosmology; indeed, the basin of
attraction of the true quintessence potential shrinks if the equation of
state today gets close to $-1$ as suggested by the observations. In this
context, it is important to check whether the coupling of the field with
other entities, gravity in this case, is able to restore the existence of
attractors for the initial dynamics of the scalar field. We find that if
the coupling is exponential the answer to that question is positive. \\ We
studied the impact of this scenario on the observations. In particular we
derived the limits from the Big Bang Nucleosynthesis; the constraints
affect both the strength of the non-minimal coupling as well as the field
value during the nuclide formation epoch. For interesting trajectories,
those bounds are largely satisfied. \\
We have finally considered the Cosmic Microwave Background (CMB) perturbation
spectra, comparing the
effects in the cases of exponential, quadratic and minimal couplings. The
main effects are due to the time
variation of the effective gravitational constant; the latter represents an
extra-source of dynamics for the Hubble expansion rate. That modifies
distances and perturbation growth rate, shifting the angular location of
the acoustic peaks in the CMB power spectrum, and enhancing its tail at low
multipoles, by means of the modified dynamics of the gravitational
potentials on large scales (Integrated Sachs-Wolfe). For a fixed
$\omega_{JBD}$ at present, a marked difference exists between the amplitude
of these effects in the two extended quintessence cosmologies considered.
Indeed, the exponential form enhances the departure from general relativity
as the field moves, with respect to the case of a quadratic coupling. The
reason is just the shape of the coupling, for which the motion of the field
gets exponentially amplified.

In conclusion, two main remarks can be made. The first one is the
capability of generalized theories of gravity to provide attractors for the
early universe dynamics of the dark energy scalar field, independently of
the shape of the potential inducing acceleration. The relevance of this
issue goes beyond the mere capability of avoiding the fine tuning on the
initial conditions. Actually, this aspects shows that the relation between
dark energy and other entities deserves a careful study, providing a rich
physics even if the present behavior is forced by observations to be close
to that of a cosmological constant. The second remark concerns the
capability of cosmology to constrain modified theories of gravity.
Interestingly the imprint on the cosmological perturbations spectra depend
on the actual form of the non-minimal interaction. The analogy existing
between different scalar-tensor theories of gravity cannot be extended to
their imprint on cosmological perturbations, which has to be done properly
case by case. This aspect is relevant having in view of the possibility to
constrain the theory of gravity on cosmological scales and from
cosmological observations, in a complementary way with respect to what is
done via observations in the solar system.

\ack We are pleased to thank M. Doran and E. Linder for useful
discussions and comments.

\end{document}